\documentclass[seceq]{ptptex}
\usepackage{wrapft}
\usepackage{graphicx}
\newcommand{\lsim}{{\stackrel{<}{\sim}}}

\def\lsim{\stackrel{\displaystyle <}{\raisebox{-1ex}{$\sim$}}}

\def\beq{\begin{eqnarray}}
\def\eeq{\end{eqnarray}}
\def\bsub{\begin{subequations}}
\def\esub{\end{subequations}}
\def\b{\begin{equation}}
\title{%        %You can use \\ for explicit line-break
On the Color-Singlet States in Many-Quark Model with the $su(4)$-Algebraic Structure. III
}
\subtitle{
Transition from the quark-triplet to the quark-pair phase
}    %use this when you want a subtitle

\author{%       %Use \sc for the family name
Yasuhiko {\sc Tsue},$^{1}$ 
Constan\c{c}a {\sc Provid\^encia},$^{2}$ 
Jo\~ao da {\sc Provid\^encia}$^{2}$ and 
Masatoshi {\sc Yamamura}$^{3}$  
}

\inst{%         %Affiliation, neglected when [addenda] or [errata]
$^{1}$Physics Division, Faculty of Science, Kochi University, Kochi 780-8520, 
Japan\\
$^{2}$Departamento de F\'{i}sica, Universidade de Coimbra, 3004-516 Coimbra, 
Portugal\\
$^{3}$Department of Pure and Applied Physics, 
Faculty of Engineering Science,\\
Kansai University, Suita 564-8680, Japan
\\

}

\recdate{%      %Editorial Office will fill in this.
\today
}

\abst{%       %this abstract is neglected when [addenda] or [errata]
A phase structure and phase transitions are investigated in the modified Bonn 
quark model. 
The force strength, which represents the particle-hole type interaction, is 
devised so as to depend on the quark number, namely quark number density. 
As a result, it is shown that the phase transition from the quark-triplet state as a nucleon to 
the quark-pair state as a color-superconducting state occurs as the density increases. 
}

\begin{document}

\maketitle

\section{Introduction}

In the series of paper, the modified Bonn quark model is investigated 
in detail about possible ground states, possible phases, order parameter and so on. 
The Bonn quark model was first introduced with a purpose of 
describing the nucleon and the $\Delta$-resonance as quark-triplet states.\cite{1}
From a modern viewpoint of studying possible phases in many-quark system or many-hadron system 
which are governed by quantum chromodynamics (QCD), namely, 
a color-superconducting phase\cite{2} and/or a quark-triplet phase as a nuclear matter, 
we reinvestigate this model under a certain extension. 
In the previous paper, in Ref.\citen{A}, which is hereafter referred to as (A), 
exact eigenstates were investigated by the method of the boson realization, 
and further, in Ref.\citen{B}, which is hereafter referred to as (B), 
the exact eigenstates were treated in a unified way. 
A phase diagram was given in Ref.\citen{C}, which is referred to as (C).  
Next, in Ref.\citen{I}, which is hereafter referred to as (I), 
the exact eigenstates are constructed so as to satisfy a certain condition which gives a ``color-singlet" state 
in average.
Based on the color-singlet state which is a color-symmetric state, the ground-state energy is 
reinvestigated in Ref.\citen{II}, which is hereafter referred to as (II).

As was mentioned in the last section of (II), the quark system at low density in the region $N^0\sim 0$ or $6\Omega$ and
at high density in the region $N^0\sim 3\Omega$ shows a quark-pair and a quark-triplet phase, respectively, 
in the original Bonn quark model with $\chi=0$. 
This situation does not change even if the fixed force strength of the particle-hole type interaction, $\chi\neq 0$, 
is simply introduced which leads to the modified Bonn quark model. 
However, the force strength $\chi$ is not fixed uniquely in this model. 
Namely, various possibilities are possible in this effective model of QCD. 
In this paper, Part III of this series, we devise the force strength $\chi$ so as to 
be suitable for the common understanding\cite{2}, that is, 
the quark-triplet phase such as nuclear matter can be realized at low quark number density and 
the quark-pair phase such as color superconducting phase can be realized at high quark number density. 
For the purpose of realizing the above-mentioned situation, we introduce a force strength $\chi$ 
which depends on the particle number. 
It is shown that, as a result, the phase transition from the quark-triplet phase at low quark number density to 
the quark-pair phase at high quark number density occurs through intermediate phases or directly.

This paper is organized as follows:
In the next section, four phases are shown and the idea of a particle-number dependent force strength 
is given. 
In Appendix A, another possibility for the particle-number dependent force strength is investigated. 
Under the force strength depending on the particle number developed in \S 2, in \S 3, 
the phase transitions and the order parameter are investigated in various patterns of $\chi$. 
In \S 4, numerical analysis is carried out and the behaviors of phase transitions are shown. 
The last section is devoted to concluding remarks.

\section{Four phases characteristics of the present model}

%%%%%%%%%%%%%%%%%%%%%%%%%%%%%%%%%%%%%%%%%%%%%%%%%%%%%%%%%%%%%%%%%%%%%%%
\begin{figure}[b]
\begin{center}
\includegraphics[height=7.5cm]{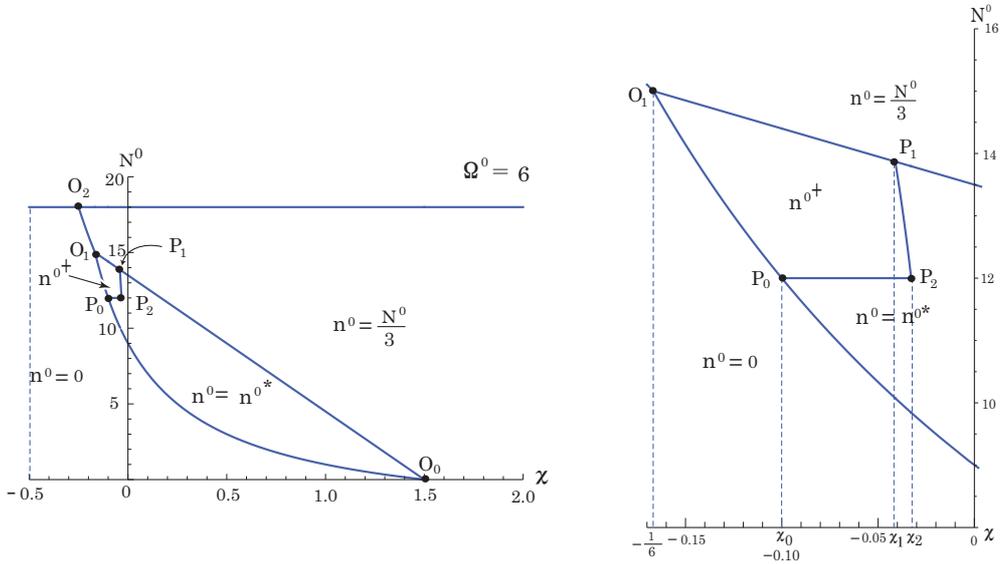}
\caption{The phase diagram is shown in the $\chi$-$N^0$ plane (left panel). The order parameter 
is regarded as $n^0$. In the right panel, the details are shown in the region of $-0.17 < \chi < 0$. 
Here, $\chi_i\ (i=0,\ 1, \ 2)$ represent $\chi_i(\Omega^0)$, respectively.}
\label{fig:2-1}
\end{center}
\end{figure}
%%%%%%%%%%%%%%%%%%%%%%%%%%%%%%%%%%%%%%%%%%%%%%%%%%%%%%%%%%%%%%%%%%%%%%%%

Our discussion starts with the interpretation of the phase diagram indicated 
in Fig.12 in (II). 
It is drawn in the $\chi$-$N^0$ plane. 
We reproduce it in Fig.{\ref{fig:2-1}}. 
The region $3\Omega^0 \leq N^0 \leq 6\Omega^0$ is symmetric with 
respect to $N^0=3\Omega^0$ to the region $0 \leq N^0 \leq 3\Omega^0$. 
Therefore, it is enough to treat the present model in the region 
$0\leq N^0 \leq 3\Omega^0$. 
Although Fig.\ref{fig:2-1} refers to the case $\Omega^0=6$, 
it illustrates the general case. 
The points characterizing the phase diagram 
are specified in terms of the following relations: 
\bsub\label{2-1}
\beq
& &{\rm O}_0\ ; \ \chi=\frac{1}{6}(2\Omega^0-3)\ , \quad N^0=0 \ , \nonumber\\
& &{\rm O}_1\ ; \ \chi=-\frac{1}{6}\ , \quad N^0=3\Omega^0-3 \ , \nonumber\\
& &{\rm O}_2\ ; \ \chi=-\frac{1}{6}\cdot\frac{\Omega^0+6}{\Omega^0+2}\ , \quad N^0=3\Omega^0 \ , 
\label{2-1a}\\
& &{\rm P}_0\ ; \ \chi=\chi_0(\Omega^0)\ , \quad N^0=2\Omega^0 \ , \nonumber\\
& &{\rm P}_1\ ; \ \chi=\chi_1(\Omega^0)\ , \quad N^0=3\Omega^0-\frac{9}{2}(1+2\chi_1(\Omega^0)) \ , \nonumber\\
& &{\rm P}_2\ ; \ \chi=\chi_2(\Omega^0)\ , \quad N^0=2\Omega^0 \ . 
\label{2-2b}
\eeq
\esub
The lines connecting these points are expressed as follows: 
\beq\label{2-2}
& &{\rm O}_0{\rm P}_1{\rm O}_1\ ; \ N^0=3\Omega^0-\frac{9}{2}(1+2\chi)\ , \nonumber\\
& &{\rm O}_0{\rm P}_0{\rm O}_1\ ; \ N^0=\frac{2\Omega^0}{1+2\chi}-3\ , \nonumber\\
& &{\rm O}_1{\rm O}_2\ ; \ N^0=\frac{6(2\Omega^0-3-6\chi)}{5+6\chi}\ , \nonumber\\
& &{\rm P}_0{\rm P}_2\ ; \ N^0=2\Omega^0\ , \nonumber\\
& &{\rm P}_1{\rm P}_2\ ; \ N^0=3\Omega^0-X_L(\chi;\Omega^0)\ . 
\eeq
We will use also the inverse relations to the relation (\ref{2-2}).

Next, we discuss the phases induced by the present model. 
The quantity $n^0$ plays the role of the order parameter. 
We can see that the present model induces four phases: 
The phase with $n^0=N^0/3$, $n^0=0$, $n^0={n^0}^*$ and $n^0={n^0}^{\dagger}$. 
The region above the line O$_0$P$_1$O$_1$O$_2$ is specified by $n^0=N^0/3$. 
We call it the quark-triplet phase. 
The region below the line O$_0$P$_0$O$_1$O$_2$ is specified by 
$n^0=0$. 
We call it the quark-pair phase. 
Two regions are surrounded by the above two phases. 
One is specified by $n^0={n^0}^*$ and the other by $n^0={n^0}^{\dagger}$. 
We call the former and the latter the intermediate phases 1 and 2, respectively. 
Hereafter, we denote the quark-triplet, the quark-pair, the intermediate phase as 
$Q_t$, $Q_p$ and $Q_i$. 
In order to distinguish 1 and 2, we use the notations $Q_{i_1}$ and $Q_{i_2}$.

For the above four phases, it may be interesting to investigate the phase transition. 
One of the ideas for this investigation is as follows: 
By changing the value of $\chi$, we trace the phase from the side 
$N^0=0$ to the side $N^0=3\Omega^0$. 
This idea was already adopted in (A) and (C), further in (II). 
On this idea, we have the following result for the tracing: 
\bsub\label{2-3}
\beq
& &(1)\ \ -\frac{1}{2} < \chi \leq -\frac{1}{6}\cdot\frac{\Omega^0+6}{\Omega^0+2}\ ; \ \ Q_p \ , 
\label{2-3a}\\
& &(2)\ \ -\frac{1}{6}\cdot\frac{\Omega^0+6}{\Omega^0+2} < \chi \leq -\frac{1}{6}\ ; \ \ Q_p \longrightarrow Q_t\ , 
\label{2-3b}\\
& &(3)\ \ -\frac{1}{6} < \chi < \frac{1}{6}(2\Omega^0-3)\ ; \ \ Q_p \longrightarrow Q_i \longrightarrow Q_t\ , 
\label{2-3c}\\
& &(4)\ \ -\frac{1}{6}(2\Omega^0-3) \leq \chi \leq \infty\ ; \ \ Q_t\ . 
\label{2-3d}
\eeq
\esub
Here, the arrow $\longrightarrow$ denotes the direction of the phase transition from the side 
$N^0=0$ to the side $N^0=3\Omega^0$. 
In the cases (1) and (4), the phase transitions do not occur. 
In the cases (2) and (3), the phase transitions occur. 
However, these phase transitions are contradictory to the common understanding. 
As was already mentioned in \S 1, the common understanding maintains that in the 
low density region $N^0 \sim 0$ and in the high density 
region $N^0 \sim 3\Omega^0$, the phases are $Q_t$ and $Q_p$, respectively. 
Our present conclusion cannot support this understanding.

In order to respond to the above situation, we must propose a new idea so as to enable us 
to give a reasonable interpretation for the common understanding. 
First, we must note that the present model does not contain any first principle 
for determining $\chi$. 
In other word, there is no necessity to adopt a fixed value of $\chi$ 
in the range $0 \leq N^0 \leq 6\Omega^0$. 
A straightforward idea may be to treat $\chi$ as an operator, which, hereafter, 
we denote as ${\hat \chi}$

For the above idea, we require that ${\hat \chi}$ is a function of ${\hat N}$ and ${\hat n}^0$. 
Of course, it depends on the $c$-number ${\Omega^0}$. 
It contains a parameter, with the help of which we can classify the phase transitions, 
for example, such as $\chi$ shown in the relation (\ref{2-3}). 
Clearly, ${\hat \chi}$ commutes with $\sum_i{\hat S}^i{\hat S}_i$ and ${\hat {\mib Q}}^2$. 
In addition to the above, we require that the results obtained in (I) and (II) are 
preserved in the present problem. 
If we follow this requirement, the eigenvalue of ${\hat \chi}$, which we denote as $\chi$, should be 
required to satisfy the condition for $\chi$, which are summarized as follows: 
\begin{description}
\item[{\rm (1)}]
\ $\displaystyle \chi > -\frac{1}{2}$, i.e., $\displaystyle 1+2\chi >0$.
\item[{\rm (2)}]
\ $\chi$ is a function of $\Omega$, $N$ and $n_0$, through the quantities $\Omega^0$ and $N^0$ 
and it is symmetric with respect to $N^0=3\Omega^0$. 
\item[{\rm (3)}]
\ In the case $N^0=0$ and $6\Omega^0$, $\chi$ should satisfy the inequality 
$\displaystyle \chi > \frac{1}{6}(2\Omega^0-3)$, i.e., $\displaystyle 
1+2\chi > \frac{2\Omega^0}{3}$. 
If $\displaystyle 1+2\chi < \frac{2\Omega^0}{3}$, it is impossible for 
the present model to be composed of quark-triplets in the region 
$N^0\sim 0$ and $6\Omega^0$. 
The above indicates that $\chi$ is sufficiently large, for example, 
$+\infty$, in the case $N^0=0$ and $6\Omega^0$. 
\end{description}
A possible candidate, which satisfies the above conditions, is as follows: 
\beq\label{2-4}
{\hat \chi}=\frac{1}{2}\left[
z\cdot \frac{3\Omega-{\hat N}}{3(\Omega-{\hat n}_0)\theta(3\Omega-{\hat N})-(3\Omega-{\hat N})+\epsilon}
-1\right] \ .
\eeq
Here, $\epsilon$ denotes a positive infinitesimal parameter and $\theta(3\Omega-{\hat N})$ is 
defined as 
\beq\label{2-5}
\theta(3\Omega-{\hat N})=\frac{3\Omega-{\hat N}+\epsilon}{\sqrt{(3\Omega-{\hat N})^2+\epsilon^2}}\ .
\eeq
Since $(3\Omega-{\hat N})^2+\epsilon^2$ is positive-definite, its square root 
is definable. 
The parameter $\epsilon$ supports that the fractional operator can be defined. 
The $c$-number $z$, which is used for the classification of the phase transition, 
satisfies 
\beq\label{2-6}
0 < z < +\infty \ .
\eeq
The operators $3\Omega-{\hat N}$ and $\Omega-{\hat n}_0$ are expressed as follows: 
\bsub\label{2-7}
\beq
& &3\Omega-{\hat N}=\frac{3}{2}({\hat b}^*{\hat b}-{\hat a}^*{\hat a}) 
+\frac{1}{2}\sum_i({\hat b}_i^*{\hat b}_i-{\hat a}_i^*{\hat a}_i) \ , 
\label{2-7a}\\
& &\Omega-{\hat n}_0=\frac{1}{2}({\hat b}^*{\hat b}+{\hat a}^*{\hat a}) 
+\frac{1}{2}\sum_i({\hat b}_i^*{\hat b}_i+{\hat a}_i^*{\hat a}_i) \ . 
\label{2-7b}
\eeq
\esub
The above two operators are taken from the relations (I$\cdot$3$\cdot$6) and 
(I$\cdot$3$\cdot$8).

Operating ${\hat \chi}$ on the eigenstates discussed in (I) and (II), 
$\chi$ is obtained in the form 
\bsub\label{2-8}
\beq
\chi=\frac{1}{2}\left[
z\cdot \frac{3\Omega^0-N^0}{3\Omega^0\theta(3\Omega^0-N^0)-(3\Omega^0-N^0)+\epsilon}-1
\right] \ , 
\label{2-8a}
\eeq
i.e., 
\beq
& &1+2\chi=z\cdot \frac{3\Omega^0-N^0}{3\Omega^0\theta(3\Omega^0-N^0)-(3\Omega^0-N^0)} \ , 
\label{2-8b}
\eeq
\esub
\beq\label{2-9}
\theta(3\Omega^0-N^0)=\frac{3\Omega^0-N^0+\epsilon}{\sqrt{(3\Omega^0-N^0)^2+\epsilon^2}} \ .
\eeq
If $3\Omega^0-N^0\neq 0$, the quantity $(3\Omega^0-N^0)$ is of the order $\epsilon^0$, 
and then, $\theta(3\Omega^0-N^0)=(3\Omega^0-N^0)/|3\Omega^0-N^0|$. 
If $3\Omega^0-N^0=0$, we have 
$\theta(3\Omega^0-N^0)=\epsilon/\sqrt{\epsilon^2}=1$. 
From the above argument, the relation (\ref{2-8b}) can be 
expressed in the form 
\beq\label{2-10}
1+2\chi=
\left\{
\begin{array}{cc}
z\left(\frac{3\Omega^0-N^0}{N^0+\epsilon}\right) \ , & (N^0 < 3\Omega^0) \\
0 \ , & (N^0=3\Omega^0) \\
z\left(\frac{N^0-3\Omega^0}{6\Omega^0-N^0+\epsilon}\right) \ . & (N^0 > 3\Omega^0) \\
\end{array}\right.
\eeq
If $N^0=0$ and $6\Omega^0$, $1+2\chi=z\cdot 3\Omega^0/\epsilon \rightarrow +\infty$ 
($\epsilon \rightarrow 0$). 
Under the above consideration, it may be permitted to express $(1+2\chi)$ in the form
\beq\label{2-11}
1+2\chi=
\left\{
\begin{array}{cc}
z\left(\frac{3\Omega^0-N^0}{N^0}\right) \ , & (0\leq N^0 \leq 3\Omega^0) \\
z\left(\frac{N^0-3\Omega^0}{6\Omega^0-N^0}\right) \ . & (3\Omega^0 \leq N^0 \leq 6\Omega^0) \\
\end{array}\right.
\eeq
The relation (\ref{2-11}) leads us 
\beq\label{2-12}
N^0=
\left\{
\begin{array}{cc}
3\Omega^0\left(1-\frac{1+2\chi}{1+2\chi+z}\right) \ , & (0\leq N^0 \leq 3\Omega^0) \\
3\Omega^0\left(1+\frac{1+2\chi}{1+2\chi+z}\right) \ . & (3\Omega^0 \leq N^0 \leq 6\Omega^0) \\
\end{array}\right.
\eeq
Hereafter, we consider the case $(0\leq N^0 \leq 3\Omega^0$). 
The quantity $N^0$ as a function of $\chi$ is monotone-decreasing from 
$N^0=3\Omega^0$ at $\chi=-1/2$ to 
$N^0\rightarrow 0$ at $\chi \rightarrow +\infty$.

The quantity $N^0$ is a function of $z$. 
Under appropriate choice of $z$, $N^0$ can pass the points P$_0$, P$_2$, P$_1$ and O$_1$. 
By searching these values of $z$, we can find the following scheme 
for the phase transition from the quark-triplet to the quark-pair: 
\bsub\label{2-13}
\beq
& &(1)\ 0<z\leq 2(1+2\chi_0(\Omega^0))\ ; \ Q_t \longrightarrow Q_{i_1} \longrightarrow Q_p \ , 
\label{2-13a}\\
& &(2)\ 2(1+2\chi_0(\Omega^0)) < z \leq 2(1+2\chi_2(\Omega^0))\ ; \ Q_t \longrightarrow Q_{i_1} \longrightarrow 
Q_{i_2}\longrightarrow Q_p \ , 
\label{2-13b}\\
& &(3)\ 2(1+2\chi_2(\Omega^0)) < z <\frac{2\Omega^0}{3}-(1+2\chi_1(\Omega^0))\ ; \ Q_t \longrightarrow Q_{i_1} \longrightarrow 
Q_{i_2}\longrightarrow Q_p \ , \nonumber\\ 
& &\label{2-13c}\\
& &(4)\ \frac{2\Omega^0}{3}-(1+2\chi_1(\Omega^0)) \leq z < \frac{2}{3}(\Omega^0-1)\ ; \ Q_t \longrightarrow 
Q_{i_2}\longrightarrow Q_p \ , 
\label{2-13d}\\
& &(5)\ \frac{2}{3}(\Omega^0-1) \leq z < +\infty \ ; \ Q_t \longrightarrow Q_p \ . 
\label{2-13e}
\eeq
\esub
Here, in the case (2) and (3), $N^0$ crosses the line P$_0$P$_2$ and the line 
P$_1$P$_2$, respectively. 
In the next section, we will discuss the above scheme in detail.

\section{Phase transitions and behavior of the order parameter}

In \S 2, we presented our basic idea for describing the $su(4)$-model 
of many-quark system. 
In this section, focusing on the phase transition from the quark-triplet to 
the quark-pair phase, we will give a concrete form of our description and 
treat the phase transitions classified in the relation (\ref{2-13}) separately. 
For the convenience of the interpretation, the behavior of the relation 
(\ref{2-12}) in the case $0\leq N^0 \leq 3\Omega^0$ is drawn on the phase diagram. 

\vspace{0.2cm}
(1) $0<z\leq 2(1+2\chi_0(\Omega^0))$ : $Q_t\rightarrow Q_{i_1}\rightarrow Q_p$. \\
We rewrite the expression of the line O$_0$P$_1$O$_1$ shown in the relation (\ref{2-2}) 
in the form 
\beq\label{3-1}
1+2\chi=\frac{2}{9}(3\Omega^0-N^0) \ .
\eeq
By equating the expression (\ref{3-1}) with the relation (\ref{2-11}) in the case 
$0\leq N^0 \leq 3\Omega^0$, we obtain the cross point of the line O$_0$P$_1$O$_1$ and the 
curve (\ref{2-12}) $(\chi_a,N_a^0)$, where $N_a^0$ is given as 
\beq\label{3-2}
N_a^0=\frac{9}{2}z \ .
\eeq
In the same way as the above, we rewrite the line O$_0$P$_0$O$_1$ shown in the relation (\ref{2-2}) :
\beq\label{3-3}
1+2\chi=\frac{2\Omega^0}{N^0+3} \ . 
\eeq
Then, the curve (\ref{2-12}) in the case $0\leq N^0 \leq 3\Omega^0$ crosses the line 
O$_0$P$_1$O$_1$ at the point 
$(\chi_b,N_b^0)$, where $N_b^0$ is given as 
\beq\label{3-4}
N_b^0=\frac{9}{2}z\cdot
\frac{4\Omega^0}{2\Omega^0-3z(\Omega^0-1)+\sqrt{4{\Omega^0}^2-12z\Omega^0(\Omega^0-1)
+9z^2(\Omega^0+1)^2}} \ .
\eeq
With the use of the relation (\ref{2-11}) in the case $0\leq N^0 \leq 3\Omega^0$, 
the order parameter $n^0$ in $Q_{i_1}$, ${n^0}^*$, is expressed in the form 
\beq\label{3-5}
{n^0}^*=
\frac{2\Omega^0N^0-z(N^0+3)(3\Omega^0-N^0)}{2N^0-3z(3\Omega^0-N^0)}
=n_c^0 \ .
\eeq
Here, we used the relation (II$\cdot$4$\cdot$8).

Thus, we can summarize the results as follows: 
\beq\label{3-6}
& &{\rm (i)}\ \ \ 0 < N^0 \leq N_a^0\ ; \qquad Q_t\ , \ n^0=\frac{N^0}{3} \ , 
\nonumber\\
& &{\rm (ii)}\ \ N_a^0 \leq N^0 \leq N_b^0\ ; \ \quad Q_{i_1}\ , \ n^0=n_c^0 \ , 
\nonumber\\
& &{\rm (iii)}\ N_b^0 \leq N^0 <3\Omega^0\ ; \quad Q_{p}\ , \ n^0=0 \ . 
\eeq
We can show that $n_c^0$ at $N^0=N_a^0$ and $N_b^0$ are equal to $N^0/3$ and 0, 
respectively. 
This means that the order parameter $n^0$ is continuous. 
In \S 4, we will show this feature explicitly. 

\vspace{0.2cm}
(2) $2(1+2\chi_0(\Omega^0)) < z \leq 2(1+2\chi_2(\Omega^0))$ : $Q_t
\rightarrow Q_{i_1}\rightarrow Q_{i_2}\rightarrow Q_p$. \\
In this case, the relations (\ref{3-2}), (\ref{3-4}) and (\ref{3-5}) are available and we 
have the following summary: 
\beq\label{3-7}
& &{\rm (i)}\ \ \ 0 < N^0 \leq N_a^0\ ; \qquad\ Q_t\ , \ n^0=\frac{N^0}{3} \ , 
\nonumber\\
& &{\rm (ii)}\ \ N_a^0 \leq N^0 < 2\Omega^0\ ; \ \quad Q_{i_1}\ , \ n^0=n_c^0 \ , 
\nonumber\\
& &{\rm (iii)}\ 2\Omega^0 < N^0 <N_b^0\ ; \quad\  Q_{i_2}\ , \ n^0=N^0-2\Omega^0 \ , 
\nonumber\\
& &{\rm (iv)}\ \ N_b^0 < N^0 < 3\Omega^0\ ; \ \quad Q_{p}\ , \ n^0=0 \ .
\eeq
It is noted that the order parameters at the transition $Q_t\rightarrow Q_{i_1}$ 
are continuous, but the transitions $Q_{i_1}\rightarrow Q_{i_2}$ and $Q_{i_2}\rightarrow Q_p$ 
are discontinuous.

\vspace{0.2cm}
(3) $\displaystyle 2(1+2\chi_2(\Omega^0)) < z < \frac{2\Omega^0}{3}-(1+2\chi_1(\Omega^0))$ : 
$Q_t \rightarrow Q_{i_1}\rightarrow Q_{i_2}\rightarrow Q_p$. \\
For this case, we must find the cross point of the line P$_1$P$_2$ given in the relation 
(II$\cdot$4$\cdot$5b) and the curve (\ref{2-12}) in the case $0\leq N^0\leq 3\Omega^0$ :
\beq\label{3-8}
1+2\chi=1+2\chi_L(3\Omega^0-N^0;\Omega^0)=z\cdot \frac{3\Omega^0-N^0}{N^0}\ .
\eeq
Since the function $\chi_L(3\Omega^0-N^0;\Omega^0)$ is of a complicated form, 
we have to adopt an approximate form. 
As can be seen in Fig.\ref{fig:2-1}, 
the line P$_1$P$_2$ seems to be close to a straight line. 
If it is permitted, we replace the line P$_1$P$_2$ by the straight line which passes two points 
$(\chi_1,N_1^0)$ and $(\chi_2,N_2^0)$:
\beq
& &1+2\chi=A-BN^0 \ , 
\label{3-9}\\
& &A=\frac{(1+2\chi_2)N_1^0-(1+2\chi_1)N_2^0}{N_1^0-N_2^0} \ , \qquad
B=\frac{(1+2\chi_2)-(1+2\chi_1)}{N_1^0-N_2^0} \ . 
\label{3-10}
\eeq
Here, $\chi_1=\chi_1(\Omega^0)$, $\chi_2=\chi_2(\Omega^0)$, 
$N_1^0=3\Omega^0-(9/2)\cdot(1+2\chi_1(\Omega^0))$ and $N_2^0=2\Omega^0$. 
Then, we can set up 
\beq\label{3-11}
1+2\chi=A-BN^0=z\cdot\frac{3\Omega^0-N^0}{N^0} \ .
\eeq
A solution satisfying $0<N^0<3\Omega^0$ is given as 
\beq\label{3-12}
N^0=\frac{6z\Omega^0}{A+z+\sqrt{A^2+2(A-6\Omega^0 B)z+z^2}}=N_c^0 \ .
\eeq

Then, we can summarize this case as follows: 
\beq\label{3-13}
& &{\rm (i)}\ \ \ 0 < N^0 \leq N_a^0\ ; \qquad Q_t\ , \ n^0=\frac{N^0}{3} \ , 
\nonumber\\
& &{\rm (ii)}\ \ N_a^0 \leq N^0 < N_c^0\ ; \ \quad Q_{i_1}\ , \ n^0=n_c^0 \ , 
\nonumber\\
& &{\rm (iii)}\ N_c^0 < N^0 <N_b^0\ ; \quad\  Q_{i_2}\ , \ n^0=N^0-2\Omega^0 \ , 
\nonumber\\
& &{\rm (iv)}\ N_b^0 < N^0 < 3\Omega^0\ ; \ \quad Q_{p}\ , \ n^0=0 \ .
\eeq
Concerning the order parameter, the transition $Q_t \rightarrow Q_{i_1}$ is 
continuous, but the others are discontinuous.

In this connection, we check the validity of the approximate form (\ref{3-9}). 
For example, in the case $\Omega^0=6$, we have 
$\chi_1(6)=-0.041099$, $\chi_2(6)=-0.032811$, 
$N_1^0=13.869891$ and $N_2^0=12$. 
Then, we obtain $A=1.040754$ and $B=0.00886469$. 
The case $N^0=13$, which is near the average of $\chi_1(6)$ and $\chi_2(6)$, gives 
us $\chi_{\rm app}=-0.037243$ and $\chi_{\rm exa}=-0.036893$. 
The error $|\chi_{\rm exa}-\chi_{\rm app}|/\chi_{\rm exa}\times 100=0.95 \%$. 
From the above case, the approximation may be rather good.

\vspace{0.2cm}
(4) $\displaystyle \frac{2\Omega^0}{3}-(1+2\chi_1(\Omega^0)) \leq z < \frac{2}{3}(\Omega^0-1)$ : 
$Q_t \rightarrow Q_{i_2}\rightarrow Q_p$. \\
Using the results already obtained, we have the following: 
\beq\label{3-14}
& &{\rm (i)}\ \ \ 0 < N^0 \leq N_a^0\ ; \qquad Q_t\ , \ n^0=\frac{N^0}{3} \ , 
\nonumber\\
& &{\rm (ii)}\ \ N_a^0 \leq N^0 < N_b^0\ ; \quad Q_{i_2}\ , \ n^0=N^0-2\Omega^0 \ , 
\nonumber\\
& &{\rm (iii)}\ N_b^0 < N^0 < 3\Omega^0\ ; \ \quad Q_{p}\ , \ n^0=0 \ .
\eeq
Two transitions are discontinuous.

\vspace{0.2cm}
(5) $\displaystyle \frac{2}{3}(\Omega^0-1) \leq z < +\infty$ : 
$Q_t \rightarrow Q_p$. \\
The line P$_1$P$_2$ is expressed as 
\beq\label{3-15}
1+2\chi=\frac{2(6\Omega^0-N^0)}{3(N^0+6)} \ .
\eeq
Then, equating the form (\ref{3-15}) with the relation (\ref{2-12}) in the case 
$0\leq N^0 \leq 3\Omega^0$, we have
\beq\label{3-16}
N^0&=&
\frac{9}{2}z\cdot \frac{8\Omega^0}{4\Omega^0-3z(\Omega^0-2)+\sqrt{16{\Omega^0}^2-24z{\Omega^0}^2
+9z^2(\Omega^0+2)^2}} \nonumber\\
&=& N_d^0 \ .
\eeq
The following is summarized: 
\beq\label{3-17}
& &{\rm (i)}\ \ \ 0 < N^0 < N_d^0\ ; \qquad Q_t\ , \ n^0=\frac{N^0}{3} \ , 
\nonumber\\
& &{\rm (ii)}\ \ N_d^0 < N^0 < 3\Omega^0\ ; \ \quad Q_{p}\ , \ n^0=0 \ .
\eeq
The transition is discontinuous.

Expression of the energy, $E^{(m)}(N^0,n^0)$, is given in the form 
\beq\label{3-18}
E^{(m)}(N^0,n^0)&=&
\frac{1}{2}z\left(\frac{3\Omega^0-N^0}{N^0}\right)
\left(\frac{1}{6}(N^0-3n^0)^2+(N^0-3n^0)\right) 
\nonumber\\
& &+\frac{1}{2}n^0(2\Omega^0-n^0)
-\frac{1}{6}N^0(6\Omega^0+6-N^0) \ . 
\ (0<N^0<3\Omega^0)\ \ \ \ 
\eeq
The form (\ref{3-18}) is derived from the relations (II$\cdot$2$\cdot$1) and 
(I$\cdot$6$\cdot$16a). 
In the next section, by showing numerical results, 
we will make 
discussion.

\section{Numerical analysis}

In this section, we present patterns of phase transitions from the quark-triplet phase at 
low density to the quark-pair phase at high density. 
The patterns of phase transitions are shown in Eq.(\ref{2-13}) and are 
discussed in detail in \S 3. Here, in this section, curves which govern the 
change of the force strength $\chi$ depending on the particle number $N^0$ are depicted 
on the phase diagram. 
Also, the ground-state energy and the order parameter are shown for a fixed value of $z$, 
which determines the behavior of the force strength $\chi$ with respect to the 
quark number.  
In this section, we adopt $\Omega^0=6$. 

%%%%%%%%%%%%%%%%%%%%%%%%%%%%%%%%%%%%%%%%%%%%%%%%%%%%%%%%%%%%%%%%%%%%%%%
\begin{figure}[t]
\begin{center}
\includegraphics[height=5.5cm]{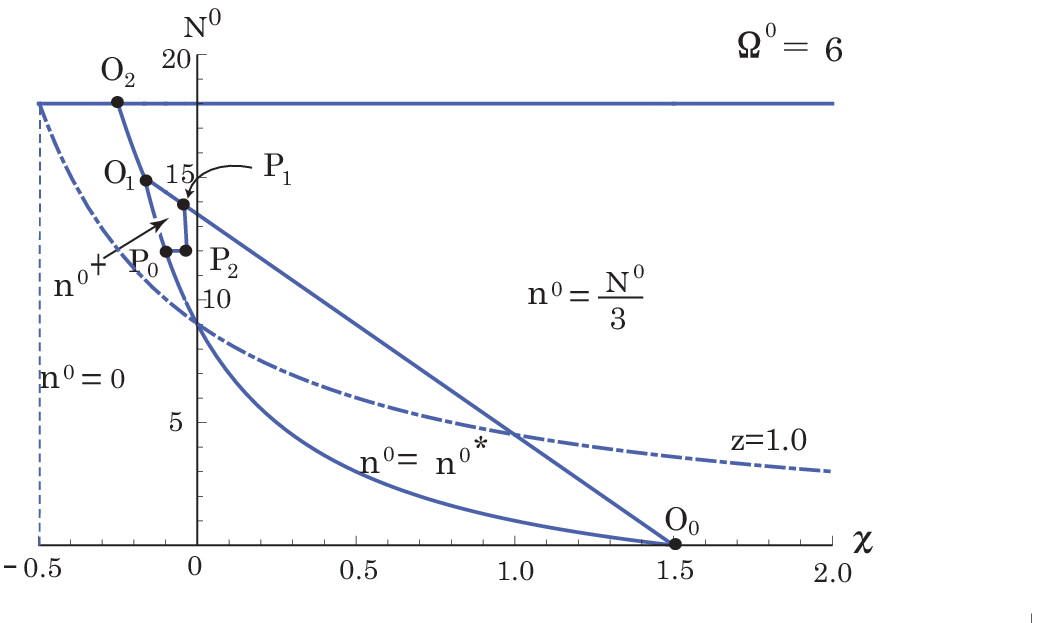}
\caption{The phase transition from quark-pair phase ($n^0=0$) at high density region 
($N^0\sim 3\Omega^0$) to quark-triplet phase ($n^0=N^0/3$) at low density region 
($N^0\sim 0$) is depicted.  
The phase transition occurs along with the dot-dashed curve which represents 
the force strength depending on the particle number $N^0$. 
The parameter $z$ is adopted to 1.0, namely, case (1) in \S 3.  
}
\label{fig:4-1}
\end{center}
\end{figure}
%%%%%%%%%%%%%%%%%%%%%%%%%%%%%%%%%%%%%%%%%%%%%%%%%%%%%%%%%%%%%%%%%%%%%%%%
%%%%%%%%%%%%%%%%%%%%%%%%%%%%%%%%%%%%%%%%%%%%%%%%%%%%%%%%%%%%%%%%%%%%%%%
\begin{figure}[t]
\begin{center}
\includegraphics[height=4.5cm]{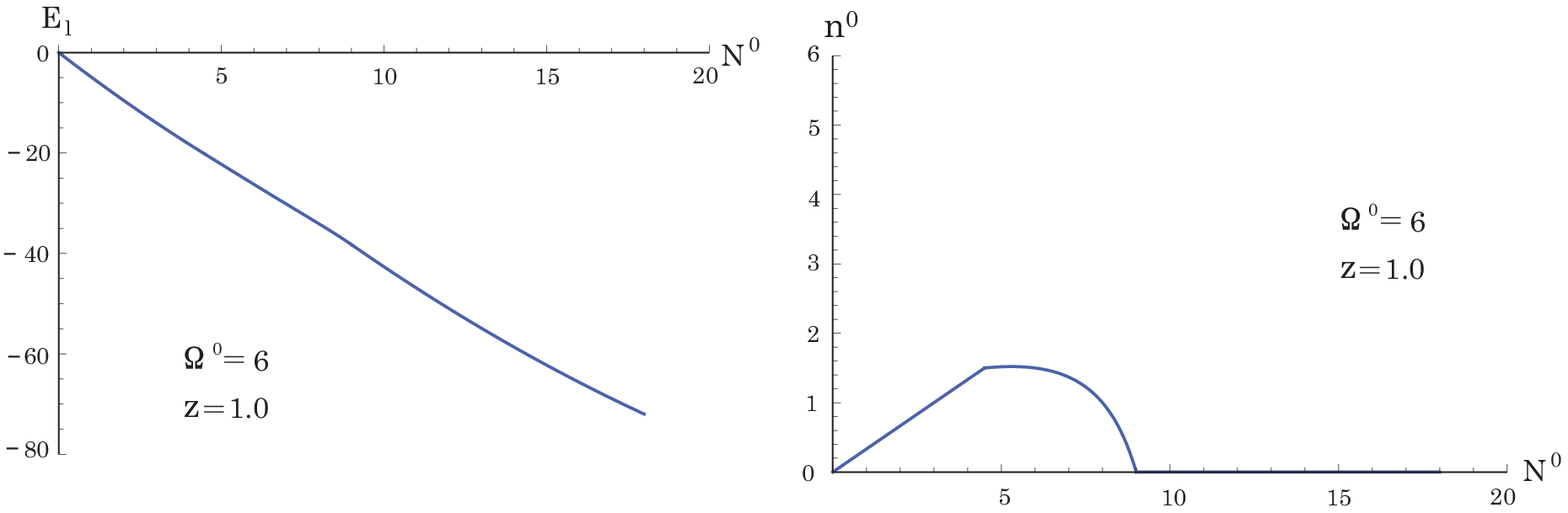}
\caption{The ground-state energy (left panel) and the order parameter (right panel) are shown 
as a function of $N^0$. 
The parameter $z$ is adopted to 1.0, namely, case (1) in \S 3.  
}
\label{fig:z1}
\end{center}
\end{figure}
%%%%%%%%%%%%%%%%%%%%%%%%%%%%%%%%%%%%%%%%%%%%%%%%%%%%%%%%%%%%%%%%%%%%%%%%
%%%%%%%%%%%%%%%%%%%%%%%%%%%%%%%%%%%%%%%%%%%%%%%%%%%%%%%%%%%%%%%%%%%%%%%
\begin{figure}[t]
\begin{center}
\includegraphics[height=7.5cm]{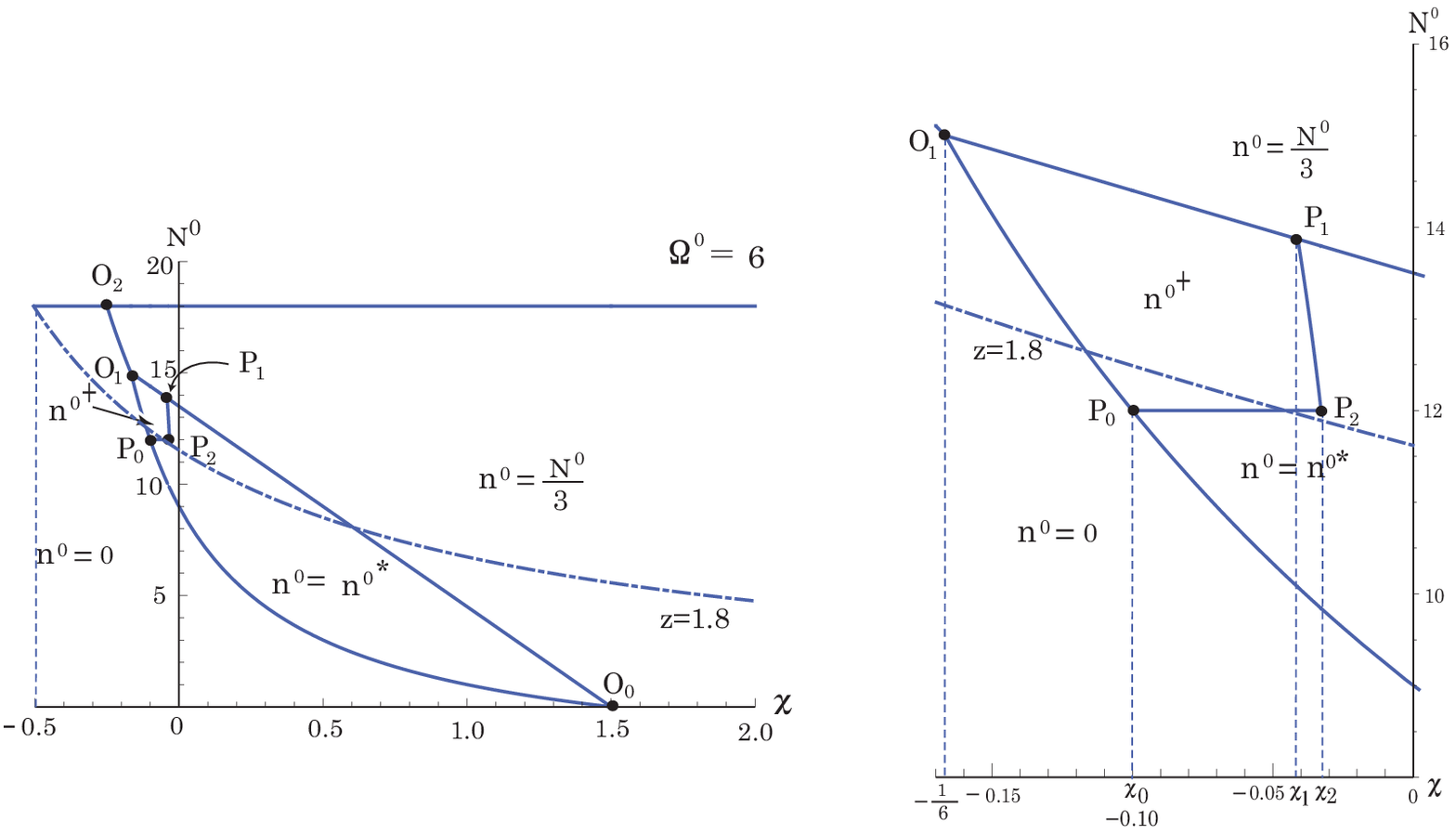}
\caption{The same as Fig.{\ref{fig:4-1}} except for $z=1.8$, namely, case (2) in \S 3.
In the right panel, the details are shown in the region of $-0.17 < \chi < 0$.  
}
\label{fig:4-2}
\end{center}
\end{figure}
%%%%%%%%%%%%%%%%%%%%%%%%%%%%%%%%%%%%%%%%%%%%%%%%%%%%%%%%%%%%%%%%%%%%%%%%
%%%%%%%%%%%%%%%%%%%%%%%%%%%%%%%%%%%%%%%%%%%%%%%%%%%%%%%%%%%%%%%%%%%%%%%
\begin{figure}[t]
\begin{center}
\includegraphics[height=4.5cm]{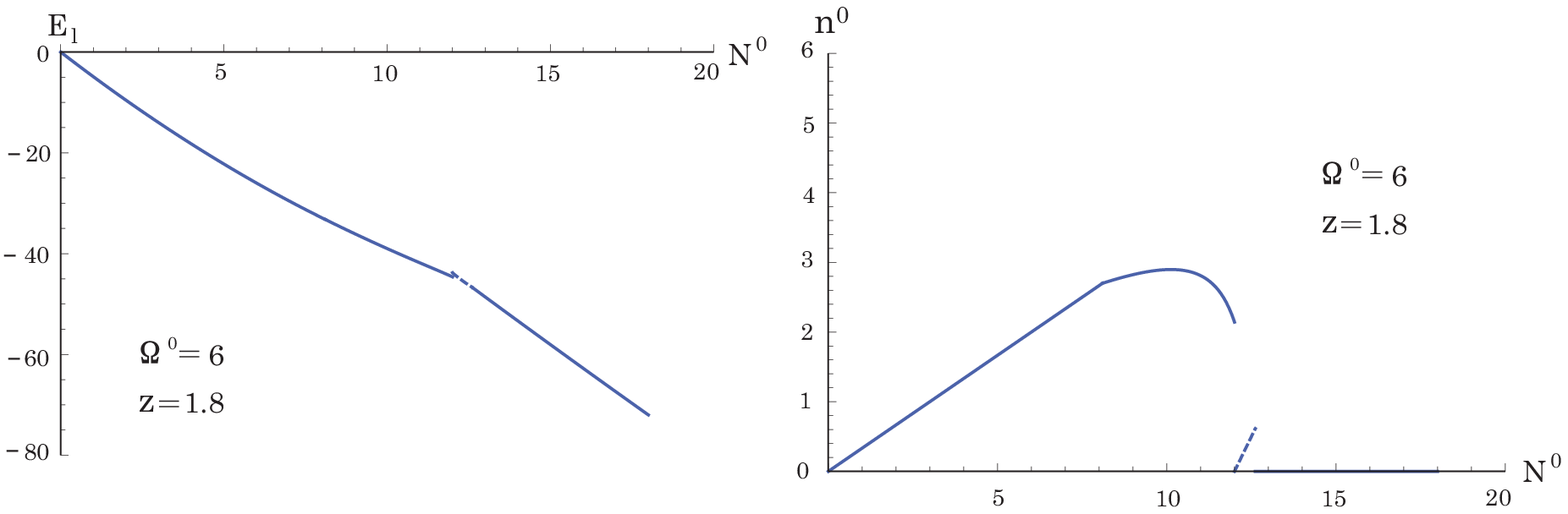}
\caption{The ground-state energy (left panel) and the order parameter (right panel) are shown 
as a function of $N^0$. 
The parameter $z$ is adopted to 1.8, namely, case (2) in \S 3.  
}
\label{fig:z2}
\end{center}
\end{figure}
%%%%%%%%%%%%%%%%%%%%%%%%%%%%%%%%%%%%%%%%%%%%%%%%%%%%%%%%%%%%%%%%%%%%%%%%

%%%%%%%%%%%%%%%%%%%%%%%%%%%%%%%%%%%%%%%%%%%%%%%%%%%%%%%%%%%%%%%%%%%%%%%
\begin{figure}[t]
\begin{center}
\includegraphics[height=7.5cm]{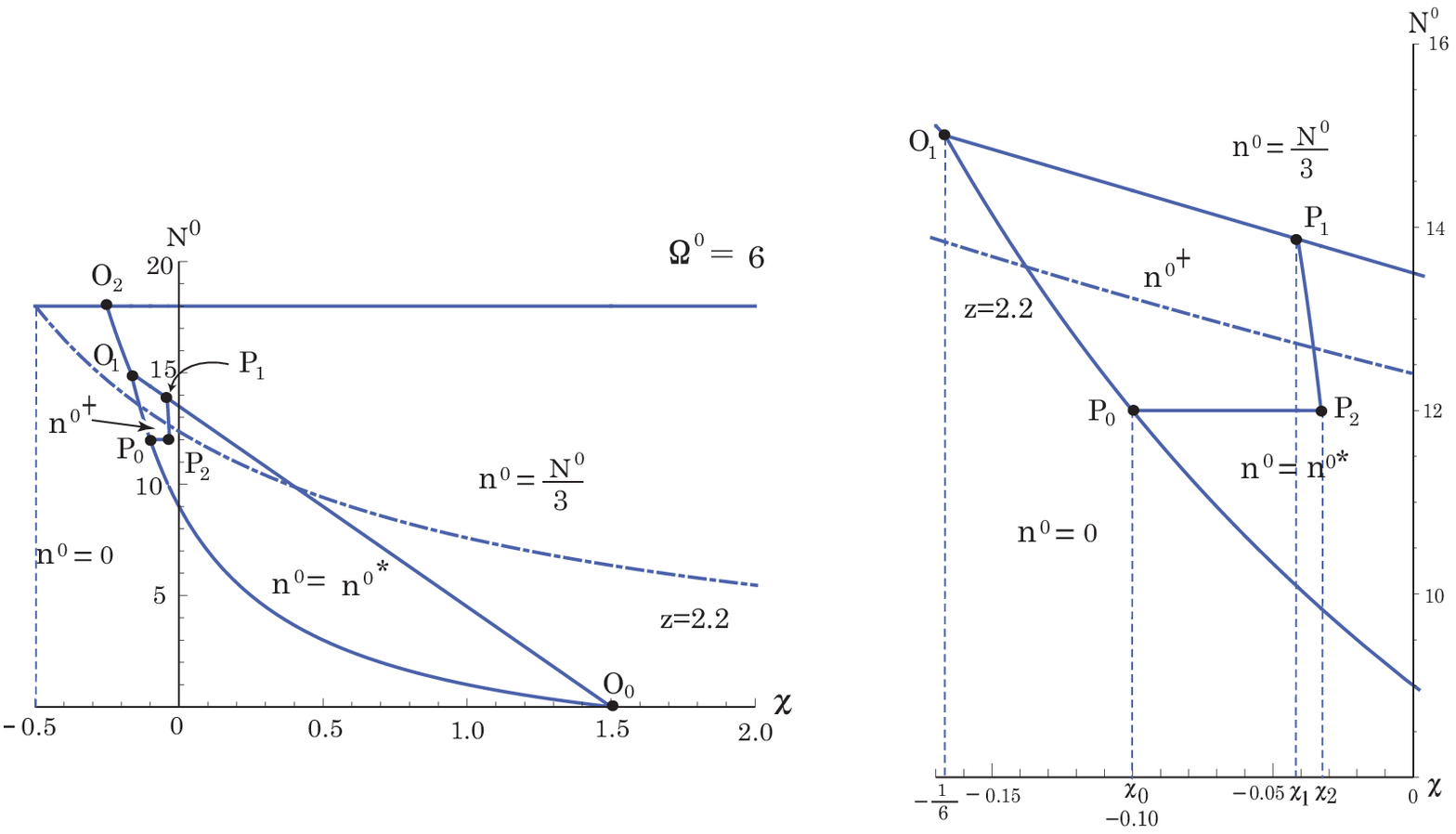}
\caption{The same as Fig.{\ref{fig:4-1}} except for $z=2.2$, namely, case (3) in \S 3.
In the right panel, the details are shown in the region of $-0.17 < \chi < 0$.  
}
\label{fig:4-3}
\end{center}
\end{figure}
%%%%%%%%%%%%%%%%%%%%%%%%%%%%%%%%%%%%%%%%%%%%%%%%%%%%%%%%%%%%%%%%%%%%%%%%
%%%%%%%%%%%%%%%%%%%%%%%%%%%%%%%%%%%%%%%%%%%%%%%%%%%%%%%%%%%%%%%%%%%%%%%
\begin{figure}[t]
\begin{center}
\includegraphics[height=4.5cm]{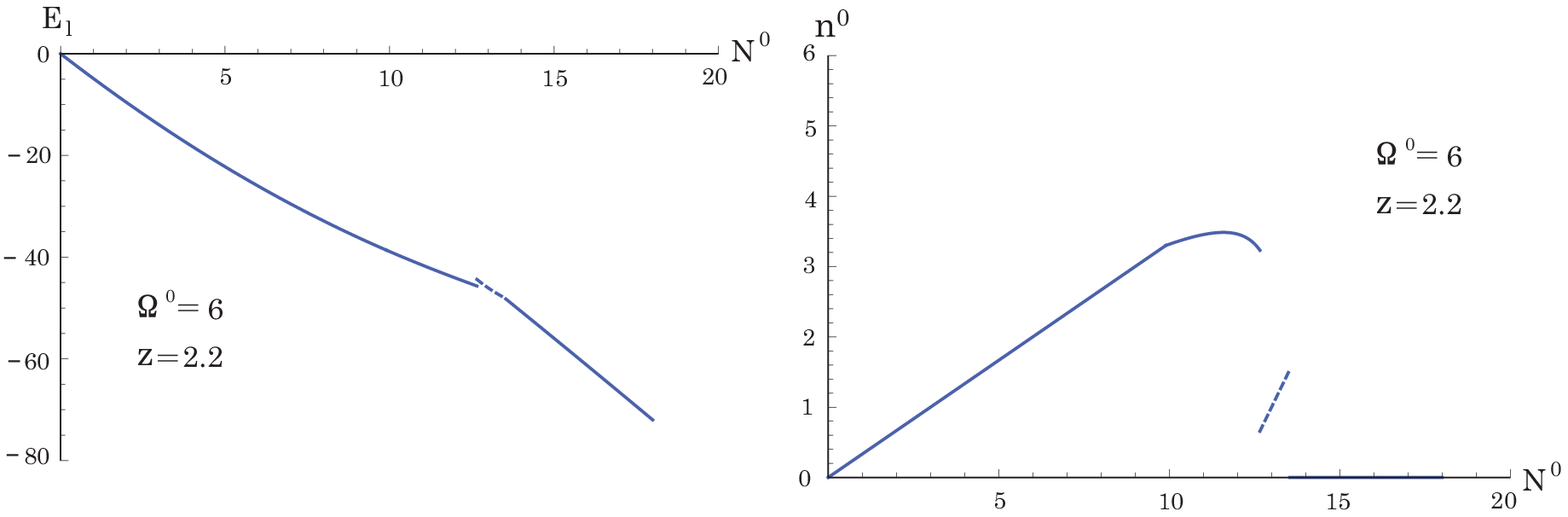}
\caption{The ground-state energy (left panel) and the order parameter (right panel) are shown 
as a function of $N^0$. 
The parameter $z$ is adopted to 2.2, namely, case (3) in \S 3.  
}
\label{fig:z3}
\end{center}
\end{figure}
%%%%%%%%%%%%%%%%%%%%%%%%%%%%%%%%%%%%%%%%%%%%%%%%%%%%%%%%%%%%%%%%%%%%%%%%

%%%%%%%%%%%%%%%%%%%%%%%%%%%%%%%%%%%%%%%%%%%%%%%%%%%%%%%%%%%%%%%%%%%%%%%
\begin{figure}[t]
\begin{center}
\includegraphics[height=7.5cm]{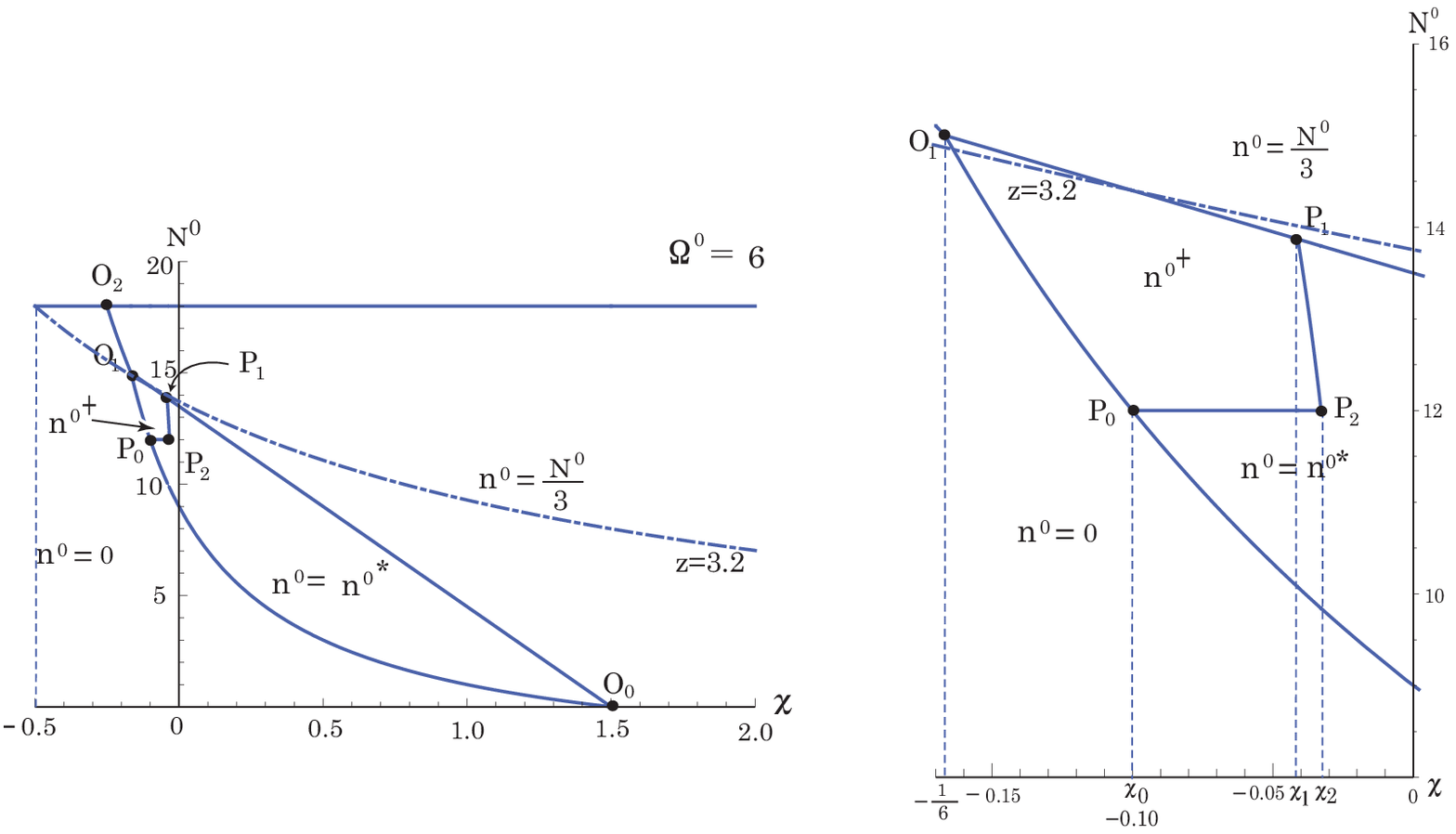}
\caption{The same as Fig.{\ref{fig:4-1}} except for $z=3.2$, namely, case (4) in \S 3.
In the right panel, the details are shown in the region of $-0.17 < \chi < 0$.  
}
\label{fig:4-4}
\end{center}
\end{figure}
%%%%%%%%%%%%%%%%%%%%%%%%%%%%%%%%%%%%%%%%%%%%%%%%%%%%%%%%%%%%%%%%%%%%%%%%
%%%%%%%%%%%%%%%%%%%%%%%%%%%%%%%%%%%%%%%%%%%%%%%%%%%%%%%%%%%%%%%%%%%%%%%
\begin{figure}[t]
\begin{center}
\includegraphics[height=4.5cm]{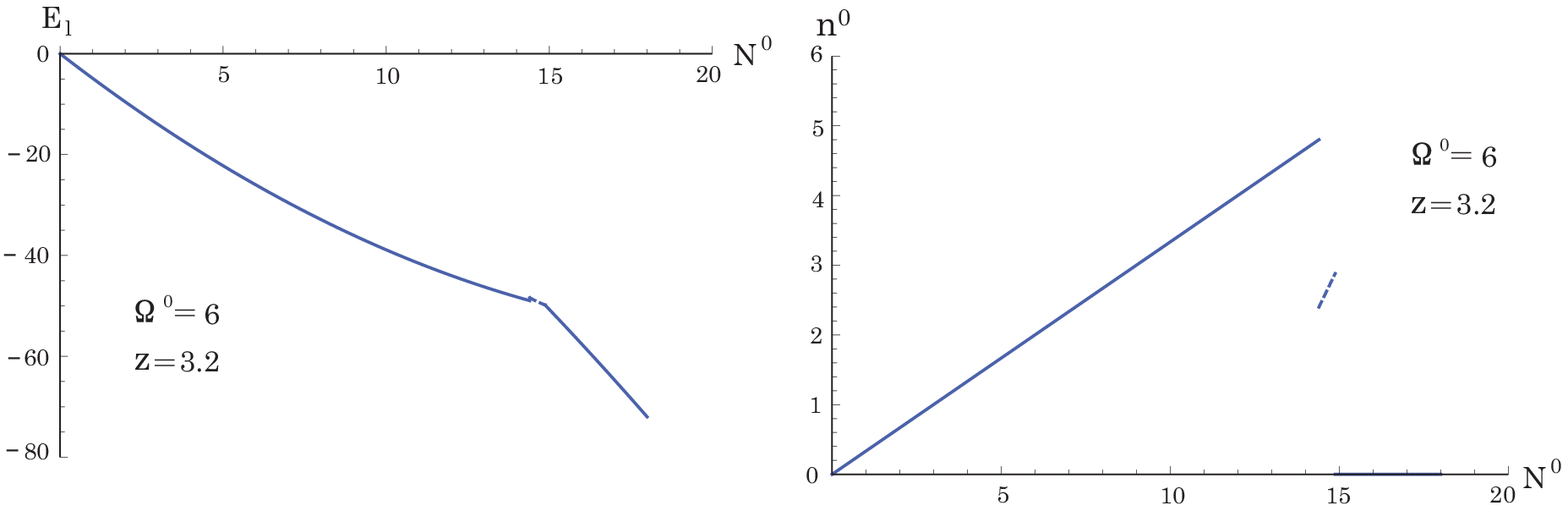}
\caption{The ground-state energy (left panel) and the order parameter (right panel) are shown 
as a function of $N^0$. 
The parameter $z$ is adopted to 3.2, namely, case (4) in \S 3.  
}
\label{fig:z4}
\end{center}
\end{figure}
%%%%%%%%%%%%%%%%%%%%%%%%%%%%%%%%%%%%%%%%%%%%%%%%%%%%%%%%%%%%%%%%%%%%%%%%

In Fig.\ref{fig:4-1}, the behavior of the phase transitions is shown in 
the case (1) in Eq.(\ref{2-13}) or in \S 3. 
Also, in Fig.\ref{fig:z1}, the ground-state energy (left panel) and the order parameter $n^0$ (right panel) 
are shown in the case (1). 
In these figures, we adopt a value of a parameter as $z=1.0$. 
At low density, namely $N^0 \leq 5$ under this parameter $z$, the quark-triplet phase is realized with 
the order parameter $n^0=N^0/3$. 
As the density increases, the force strength changes from large positive value to $-1/2$ monotonically.
In this case, the intermediate phase with the order parameter $n^0={n^0}^*(=n_c)$ is realized 
in a certain density region. 
After that, at high density region, namely $N^0\sim 3\Omega^0$, the force strength $\chi$ is close to $-1/2$. 
In the high density region around $N^0\sim 3\Omega^0$, the quark-pair phase is realized with the order parameter 
$n^0=0$. 
In this case, the order parameter changes continuously in both phase transitions.

In Fig.\ref{fig:4-2}, the behavior of the phase transitions is shown in 
the case (2) in Eq.(\ref{2-13}) or in \S 3. 
Also, in Fig.\ref{fig:z2}, the ground-state energy (left panel) and the order parameter $n^0$ (right panel) 
are shown in the case (2). 
In these figures, we adopt a value of a parameter as $z=1.8$. 
In the low density region, the quark-triplet phase is realized with the order parameter $n^0=N^0/3$. 
As the density increases, first, the intermediate phase is realized with the order parameter 
$n^0={n^0}^{*}(=n_c)$. Secondly, another intermediate phase is realized with $n^0={n^0}^{\dagger}$. 
Then, the curve which governs the force strength depending on $N^0$ crosses the curve P$_0$P$_2$. 
The details are depicted in the right panel in Fig.\ref{fig:4-2}. 
In the transition from the intermediate phase $Q_{i_1}$ with the order parameter $n^0=n_c$ to 
the intermediate phase $Q_{i_2}$ with $n^0=n^{0\dagger}$, the order parameter and the ground-state energy are 
changed discontinuously. 
In the transition from the intermediate phase $Q_{i_2}$ with $n^0=n^{0\dagger}$ 
to the quark-pair phase with $n^0=0$, the order parameter is changed discontinuously, but 
the ground-state energy is changed continuously.  
In these figures, the region with $n^0=n^{0\dagger}$ is indicated by dotted curve.
Finally, in the high density region, the quark-pair phase is realized with the order parameter 
$n^0=0$.

In Fig.\ref{fig:4-3}, the behavior of the phase transitions is shown in 
the case (3) in Eq.(\ref{2-13}) or in \S 3. 
Also, in Fig.\ref{fig:z3}, the ground-state energy (left panel) and the order parameter $n^0$ (right panel) 
are shown in the case (3). 
In these figures, we adopt a value of a parameter as $z=2.2$. 
The behaviors of the phase transitions are the same as those in Fig.\ref{fig:4-2}, except that 
the curve which determines the force strength depending on $N^0$ crosses the curve P$_1$P$_2$. 
The details are depicted in the right panel in Fig.\ref{fig:4-3}. 
In the transition from the phase with the order parameter $n^0=N^0/3$ to one with $n^0=n^{0\dagger}$, 
the order parameter and the ground state energy are changed discontinuously. 
In the transition from the intermediate phase  with $n^0=n^{0\dagger}$ 
to the quark-pair phase with $n^0=0$, the order parameter is changed discontinuously, but 
the ground-state energy is changed continuously. 
These behaviors are similar to those seen in the case (2).   
In these figures, the region with $n^0=n^{0\dagger}$ is indicated by dotted curve.

In the case $z=3.2$, which leads to the case (4) in Eq.(\ref{2-13}) or in \S 3, 
the phase transition occurs from the quark-triplet phase with $n^0=N^0/3$ to 
the intermediate phase with $n^0={n^0}^{\dagger}$, as the density increases. 
After that, the quark-pair phase is realized. 
These behaviors are shown in Figs.\ref{fig:4-4} and {\ref{fig:z4}}. 
In the transition from the phase with the order parameter $n^0=N^0/3$ to one with $n^0=n^{0\dagger}$, 
the order parameter and the ground state energy are changed discontinuously. 
In the transition from the intermediate phase  with $n^0=n^{0\dagger}$ 
to the quark-pair phase with $n^0=0$, the order parameter is changed discontinuously, but 
the ground-state energy is changed continuously. 
In these figures, the region with $n^0=n^{0\dagger}$ is indicated by dotted curve.

%%%%%%%%%%%%%%%%%%%%%%%%%%%%%%%%%%%%%%%%%%%%%%%%%%%%%%%%%%%%%%%%%%%%%%%
\begin{figure}[t]
\begin{center}
\includegraphics[height=5cm]{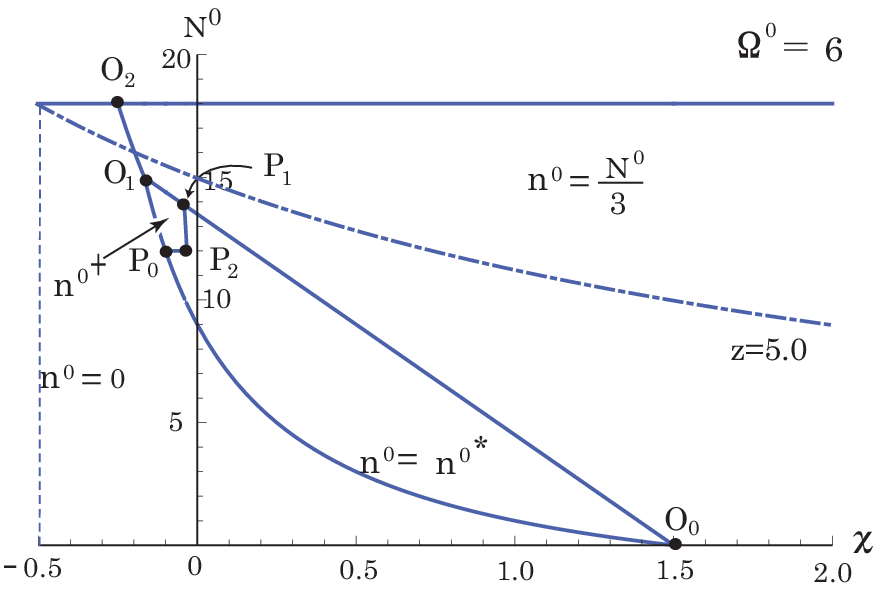}
\caption{The same as Fig{\ref{fig:4-1}} except for $z=5.0$, namely, case (5) in \S 3.
}
\label{fig:4-5}
\end{center}
\end{figure}
%%%%%%%%%%%%%%%%%%%%%%%%%%%%%%%%%%%%%%%%%%%%%%%%%%%%%%%%%%%%%%%%%%%%%%%%
%%%%%%%%%%%%%%%%%%%%%%%%%%%%%%%%%%%%%%%%%%%%%%%%%%%%%%%%%%%%%%%%%%%%%%%
\begin{figure}[t]
\begin{center}
\includegraphics[height=4.5cm]{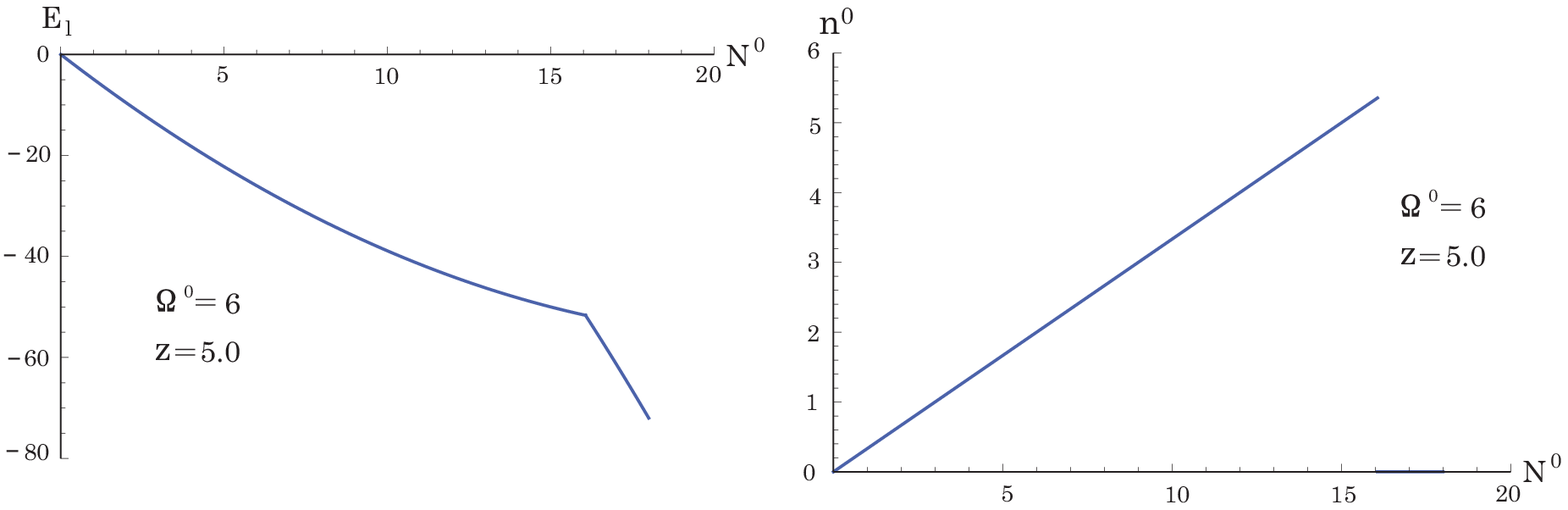}
\caption{The ground-state energy (left panel) and the order parameter (right panel) are shown 
as a function of $N^0$. 
The parameter $z$ is adopted to 5.0, namely, case (5) in \S 3.  
}
\label{fig:z5}
\end{center}
\end{figure}
%%%%%%%%%%%%%%%%%%%%%%%%%%%%%%%%%%%%%%%%%%%%%%%%%%%%%%%%%%%%%%%%%%%%%%%%

Finally, in Fig.\ref{fig:4-5}, the behavior of the phase transition is shown in 
the case (5) in Eq.(\ref{2-13}) or in \S 3.
The parameter $z$ is taken as 5.0. 
In this case, the phase transition occurs directly from the quark-triplet phase at low density to 
the quark-pair phase at high density. 
The order parameter is changed discontinuously, but the ground-state energy is changed continuously.

Originally, the parameters $N^0$ and $n^0$ are positive integers. 
It should be noted that the numerical results shown in this section 
are based on the idea in which $N^0$ and $n^0$ are continuously changing parameters.

%\newpage
\section{Concluding remarks}

In this series of papers, the modified Bonn quark model was widely investigated 
from the viewpoint of the possible phases and phase transitions. 
The quark-triplet state which is regarded as nucleon is included in the Bonn quark model. 
Further, the quark-pair state which is regarded as the color superconducting state is 
also included. 
Thus, the realized phase and the phase transition between the nucleon state and the color pairing state 
can be studied in this modified Bonn quark model. 
In the first paper of this series, namely Part I,\cite{I} 
the color-singlet states were constructed as the color-symmetric states by means of the 
color-singlet condition and the minimization condition of the expectation value of the 
$su(3)$-Casimir operator.  
This construction was carried out in the boson space by using of the boson realization method. 
In the second paper of this series, namely Part II,\cite{II} 
The ground-state energies were calculated and the ground state was determined for each 
force strength $\chi$. 
Then, it was shown that the variable $n^0$ plays the role of an order parameter. 

In this paper, namely Part III, the phase structure on $\chi$-$N^0$ plane and the 
phase transitions are investigated. 
It was shown that there were four phases which are the quark-triplet phase, quark-pair phase and 
two intermediate phases. 
The order parameters were $n^0=N^0/3$, $n^0=0$ and $n^0\equiv n^{\dagger}=N^0-2\Omega^0$ and 
$n^0=n_c^0$, respectively. 
In this paper, the force strength depending on the quark number were introduced, and then, 
it was shown that the phase transition occurs from the quark-triplet state at low density to 
the quark-pair state at high density in various paths on the $\chi$-$N^0$ plane. 
Also, the behavior of the order parameter was investigated where the order parameter 
changes continuously or discontinuously through the phase transition.  
Further, another possibility of the path of the phase transition on $\chi$-$N^0$ plane 
was discussed.

With the use of the operator ${\hat \chi}$ defined in the relation (\ref{2-4}), 
we discussed the phase transitions appearing in the present model. 
For a given value of $z$, $N_a^0$ and $N_b^0$ are uniquely fixed. 
For example, in the case $\Omega^0=6$, $z=1.6$ fixes $N_a^0$ and $N_b^0$ 
to be 7.2 and 12, respectively. 
This case passes through the point P$_0$. 
The above result tells us that the quark-numbers in $Q_t$, $Q_i$ and $Q_p$ are 
roughly 7, 5 and 6. 
Including the results shown in Figs.\ref{fig:4-1}$\sim$\ref{fig:z5}, 
this fact gives us the following impression:
In the case $z\lsim 1.6$, we observe that the phase transition from $Q_t$ to 
$Q_p$ seems to occur after rather large change of the quark-number. 
If we expect the transition after smaller changes of the quark-number, 
further consideration may be necessary. 
In the Appendix, a basic idea for this problem will be sketched. 
Of course, the results obtained in this paper are altered quantitatively, 
but, qualitatively, almost all the results are unchanged. 
Therefore, basic part of our idea presented in this paper is conserved and the 
treatment in the Appendix may be helpful for more thorough investigation 
on the $su(4)$-model of many-quark system.

%\section*{Acknowledgements} 

\appendix
\section{A possible generalization of the form (\ref{2-4})}

In this Appendix, we will present a possible generalization of the 
form (\ref{2-4}), with the aid of which we can discuss the problem mentioned 
in the ending of \S 5. 
In addition to $z$, we introduce a real parameter $\alpha$ and 
define the following form:
\beq\label{a1}
{\hat \chi}=\frac{1}{2}
\left[z\cdot\frac{3\Omega-{\hat N}}{3(\Omega-{\hat n}_0)(\theta(3\Omega-{\hat N})-\alpha)
-(3\Omega-{\hat N})+\epsilon}-1\right] \ . 
\eeq
Operating ${\hat \chi}$ on the eigenstate of ${\hat N}$ and ${\hat n}_0$ with 
the same idea as that in the case (\ref{2-11}), we have 
\beq\label{a2}
1+2\chi=z\left(\frac{3\Omega^0-N^0}{N^0-3\Omega^0\alpha}\right) \ . \quad
(0\leq N^0 \leq 3\Omega^0)
\eeq
Inversely, the relation (\ref{a2}) is rewritten as 
\beq\label{a3}
N^0=3\Omega^0\left(1-\frac{(1+2\chi)(1-\alpha)}{1+2\chi+z}\right) \ . 
\quad (0\leq N^0 \leq 3\Omega^0)
\eeq
We can see that if $\alpha=0$, the forms (\ref{a2}) and (\ref{a3}) are 
reduced to the forms (\ref{2-11}) and (\ref{2-12}), respectively.

\begin{figure}[t]
\begin{center}
\includegraphics[height=3.1cm]{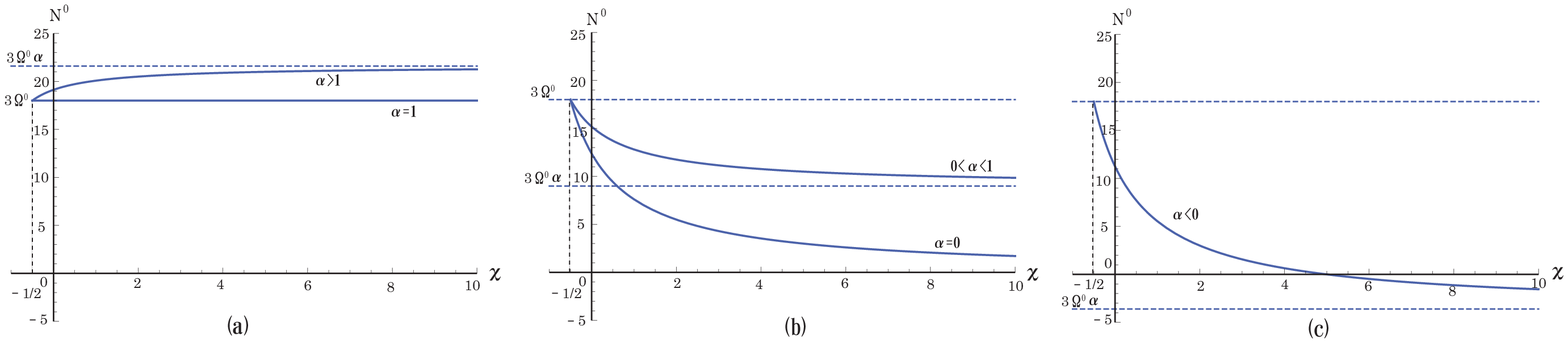}
\caption{The relation between $N^0$ and $\chi$ is shown in various $\alpha$. 
(a) $\alpha=1$ and $\alpha (=1.2)>1$. (b) $\alpha=0$ and $0<\alpha (=0.5) <1$. 
(c) $\alpha (=-0.2) <0$.
}
\label{fig:X}
\end{center}
\end{figure}

\begin{figure}[t]
\begin{center}
\includegraphics[height=4.5cm]{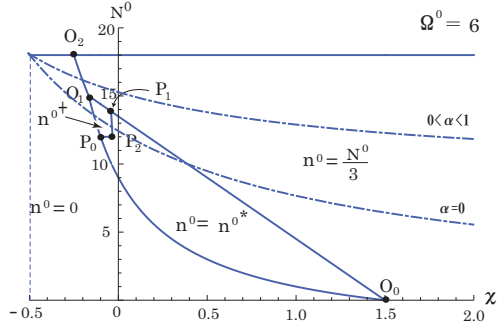}
\caption{The path of phase transition is shown in each case of $\alpha=0$ and $0<\alpha<1$ 
in the case $z=2.2$. 
Here, we adopt $\alpha=0.5$ for the case $0<\alpha<1$. 
}
\label{fig:phase-alpha}
\end{center}
\end{figure}

We draw the relation (\ref{a3}) on the $\chi$-$N^0$ plane. 
As is clear from the behavior shown in Fig.{\ref{fig:X}}(a), the case 
$\alpha \geq 1$ may be meaningless for the present model. 
With the help of the ``color-singlet" state, we can determine the lowest limit of $\chi$, 
i.e., $\chi=-1/2$. 
But, we do not have any condition to determine the upper limit of $\chi$. 
It suggests that we must consider the present model in the range 
$-1/2<\chi < +\infty$. 
If there exists a finite upper limit of $\chi$, the case $\alpha <0$ 
shown in Fig.\ref{fig:X}(c) may be interesting. 
But, the existence of the upper limit of $\chi$ is denied, and then, the case 
$\alpha<0$ may be also not interesting for the present model. 
From the above observation, our central interest is in the case 
$0\leq \alpha <1$ shown in Fig.\ref{fig:X}(b). 
As was already mentioned, the case $\alpha=0$ was treated in this paper 
in detail and at the limit $\chi\rightarrow +\infty$, 
$N^0\rightarrow 0$. 
In the case $0<\alpha <1$, at the limit $\chi\rightarrow +\infty$, 
$N^0\rightarrow 3\Omega^0\alpha$. 
Therefore, the case $0<\alpha <1$ can be regarded as an extension of the 
condition (3) developed in \S 2. 
The path of phase transition is shown in Fig.\ref{fig:phase-alpha}.

%%%%%%%%%%%%%%%%%%%%%%%%%%%%%%%%%%%%%%%%%%%%%%%%%%%%%%%%%%%%%%%%%%%%%%%
%%%%%%%%%%%%%%%%%%%%%%%%%%%%%%%%%%%%%%%%%%%%%%%%%%%%%%%%%%%%%%%%%%%%%%%%

%%%%%%%%%%%%%%%%%%%%%%%%%%%%%%%%%%%%%%%%%%%%%%%%%%%%%%%%%%%%%%%%%%%%%%%
%%%%%%%%%%%%%%%%%%%%%%%%%%%%%%%%%%%%%%%%%%%%%%%%%%%%%%%%%%%%%%%%%%%%%%%%

In \S 3, we discussed the phase transitions from $Q_t$ to $Q_p$ by 
classifying the behaviors of the relation (\ref{2-12}) for the case 
$0\leq N^0 \leq 3\Omega^0$ under the five cases shown in the relation (\ref{2-13}). 
After rather lengthy consideration, we can generalize the relation (\ref{2-12}) as 
follows:
\bsub\label{a4}
\beq
& &(\alpha_1)\ \ 0\leq \alpha <\frac{2}{3} \ , 
\label{a4a}\\
& &\quad (1)\quad 0< z \leq (2-3\alpha)(1+2\chi_0(\Omega^0)) \ , \nonumber\\
& &\quad (2)\quad (2-3\alpha)(1+2\chi_0(\Omega^0)) < z \leq (2-3\alpha)(1+2\chi_2(\Omega^0)) \ , 
\nonumber\\
& &\quad (3)\quad (2-3\alpha)(1+2\chi_2(\Omega^0))<z <\frac{2}{3}\Omega^0(1-\alpha)-(1+2\chi_1(\Omega^0)) \ , 
\nonumber\\
& &\quad (4)\quad \frac{2}{3}\Omega^0(1-\alpha)-(1+2\chi_1(\Omega^0))
\leq z <\frac{2}{3}(\Omega^0(1-\alpha)-1) \ , 
\nonumber\\
& &\quad (5)\quad \frac{2}{3}(\Omega^0(1-\alpha)-1) \leq z < +\infty \ , 
\nonumber\\
& &(\alpha_2)\ \ \frac{2}{3} \leq \alpha < 1-\frac{3}{2}\frac{1+2\chi_1(\Omega^0)}{\Omega^0} \ ,  
\label{a4b}\\
& &\quad (3)\quad 0< z \leq\frac{2}{3}\Omega^0(1-\alpha) \ , \nonumber\\
& &\quad (4)\quad \frac{2}{3}\Omega^0(1-\alpha)-(1+2\chi_1(\Omega^0))
\leq z <\frac{2}{3}(\Omega^0(1-\alpha)-1) \ , 
\nonumber\\
& &\quad (5)\quad \frac{2}{3}(\Omega^0(1-\alpha)-1) \leq z < +\infty \ , 
\nonumber\\
& &(\alpha_3)\ \ 1-\frac{3}{2}\frac{1+2\chi_1(\Omega^0)}{\Omega^0} \leq \alpha < 1-\frac{1}{\Omega^0}\ ,  
\label{a4c}\\
& &\quad (4)\quad 0 < z <\frac{2}{3}(\Omega^0(1-\alpha)-1) \ , 
\nonumber\\
& &\quad (5)\quad \frac{2}{3}(\Omega^0(1-\alpha)-1) \leq z < +\infty \ , 
\nonumber\\
& &(\alpha_4)\ \ 1-\frac{1}{\Omega^0} \leq \alpha < 1 \ ,  
\label{a4d}\\
& &\quad (5)\quad 0 < z < +\infty \ . \nonumber 
\eeq
\esub
Here, the numbering (1) $\sim$ (5) corresponds to the numbering 
(1) $\sim$ (5) shown in the relation (\ref{2-13}). 
On the basis of the relations (\ref{a2}) and (\ref{a3}), together with the relation (\ref{a4}), 
we can describe the phase transitions appearing in the present model. 
But, the full description is nothing but the repetition.

Finally, we will contact with the problem mentioned in the ending of \S 5, 
which gives us the motivation of the present generalization. 
In the range $0 < N^0 < 3(\Omega^0-1)$, $N_a^0$ and $N_b^0$ satisfy the relations 
\bsub\label{a5}
\beq
& &\frac{2}{9}(3\Omega^0-N_a^0)=z\left(\frac{3\Omega^0-N_a^0}{N_a^0-3\Omega^0\alpha}\right)
\ (=1+2\chi_a) \ , 
\label{a5a}\\
& &\frac{2\Omega^0}{N_b^0+3}=z\left(\frac{3\Omega^0-N_b^0}{N_b^0-3\Omega^0\alpha}\right)
\ (=1+2\chi_b) \ .
\label{a5b}
\eeq
\esub
Here, we used the relation (\ref{2-2}) for the lines O$_0$P$_1$O$_1$ and 
O$_0$P$_0$O$_1$. 
With the use of the relation (\ref{a5}), $N_a^0$ and $N_b^0$ are expressed 
in terms of $z$ and $\alpha$. 
Hereafter, we express $z$ and $N_a^0$ in terms of $\alpha$ and $N_b^0$: 
\beq
& &z=\frac{2\Omega^0(N_b^0-N_a^0)}{N_b^0(3(\Omega^0-1)-N_b^0)} \ , 
\label{a6}\\
& &N_a^0=\frac{3\Omega^0N_b^0(3+\alpha(3(\Omega^0-1)-N_b^0))}{N_b^0(3(\Omega^0-1)-N_b^0)+9\Omega^0} \ .
\label{a7}
\eeq
Here, $(N_b^0-N_a^0)$ is given as 
\beq\label{a8}
N_b^0-N_a^0=\frac{N_b^0(3(\Omega^0-1)-N_b^0)(N_b^0-3\alpha\Omega^0)}{
N_b^0(3(\Omega^0-1)-N_b^0)+9\Omega^0} \ . 
\eeq
The quantities $N_a^0$, $(N_b^0-N_a^0)$ and $(3\Omega^0-N_b^0)$ denote 
the quark-numbers in $Q_t$, $Q_i$ and $Q_p$, respectively. 
Of course, if $\alpha=0$, the above results reduce to those given in this paper. 
We can see that if $\alpha >0$, $N_a^0$ increases from that in the case 
$\alpha=0$ and $(N_b^0-N_a^0)$ decreases from that in the case $\alpha=0$. 
Therefore, we can expect that the above is a solution of our problem. 
For example, in the case $\Omega^0=6$, if $\alpha =0$, $N_a^0=7.2$, 
$N_b^0-N_a^0=2.4$ and $3\Omega^0-N_b^0=6$. 
Clearly, the phase transition from $Q_t$ to $Q_p$ occurs after smaller change of the 
quark-number $(4.8 \rightarrow 2.4)$. 
As was already mentioned, we must investigate the case $\alpha <0$, 
if there exists the finite upper limit for $\chi$. 
If $\alpha <0$, $N_a^0$ decreases and $(N_b^0-N_a^0)$ increases. 
This indicates that the transition occurs after larger change of the 
quark number. 
In the case where $|\alpha|$ is appropriately large, the quark-triplet phase becomes minor and 
the present model loses its reason for existing. 
In this sense, $\alpha$ may be under the condition $0\leq \alpha <1$.

\end{document}